\documentclass[12pt,preprint]{aastex}

\def\lesssim{\mathrel{\hbox{\rlap{\hbox{\lower4pt\hbox{$\sim$}}}\hbox{$<$}}}}
\def\gtrsim{\mathrel{\hbox{\rlap{\hbox{\lower4pt\hbox{$\sim$}}}\hbox{$>$}}}}

\usepackage{graphicx}
\input{psfig.sty}

\slugcomment{}

\title{Thermodynamics of an Accretion Disk Annulus with
Comparable Radiation and Gas Pressure}

\shortauthors{Krolik et al.}

\begin{document}

\shorttitle{Thermodynamics of Accretion Disks}

\author{Julian H. Krolik}
\affil{Department of Physics and Astronomy, Johns Hopkins University, 
    Baltimore, MD 21218}

\author{Shigenobu Hirose}
\affil{The Earth Simulator Center, JAMSTEC, Yokohama, Kanagawa 236-0001, Japan}

\and 

\author{Omer Blaes}
\affil{Department of Physics, University of California, Santa Barbara,
Santa Barbara CA 93106}



\begin{abstract}

We explore the thermodynamic and global structural
properties of a local patch of an
accretion disk whose parameters were chosen so that radiation
pressure and gas pressure would be comparable in magnitude.
Heating, radiative transport, and cooling are computed
self-consistently with the structure by solving the equations of
radiation MHD in the shearing-box approximation.  Using
a fully 3-d and energy-conserving code, we compute the structure
and energy balance of this disk segment over a span of more than forty
cooling times.  As is also true when gas pressure dominates,
the disk's upper atmosphere is magnetically-supported.  However,
unlike the gas-dominated case, no steady-state is reached; instead,
the total (i.e., radiation plus gas) energy content fluctuates by
factors of 3--4 over timescales of several tens of orbits, with no
secular trend.  Because
the radiation pressure varies much more than the gas pressure, the ratio
of radiation pressure to gas pressure varies over the range $\simeq 0.5$--2.
The volume-integrated dissipation rate generally increases with increasing
total energy, but the mean trend is somewhat slower than linear, and the
instantaneous dissipation rate
is often a factor of two larger or smaller than the mean for that total
energy level.  Locally, the dissipation rate per unit volume
scales approximately in proportion to the current density; the time-average
dissipation rate per unit mass is proportional to $m^{-1/2}$, where
$m$ is the horizontally-averaged mass column density to the nearer of
the top or bottom surface.  As in our earlier
study of a gas-dominated shearing-box, we find that energy transport is
completely dominated by radiative diffusion, with Poynting flux carrying
less than $1\%$ of the energy lost from the box.

\end{abstract}


\section{Introduction}

        In the years since the initial recognition that magneto-rotational
instability (MRI) is the key to angular momentum transport, and therefore mass inflow,
through accretion disks, simulations of accretion dynamics have become ever more
sophisticated (e.g., Hawley \& Balbus 1991; Stone et~al. 1996; Brandenburg et al.
1995; Miller \& Stone 2000; Hawley \& Krolik 2001; De Villiers et~al. 2003;
McKinney \& Gammie 2004).  However, although it has long been understood that
dissipation is inextricably tied to accretion (Novikov \& Thorne 1973) and
radiative cooling is what makes accretion astronomically interesting, nearly
all these calculations have sidestepped a crucial physical issue: disk
thermodynamics.  

        This neglect has numerous
consequences: The temperature
of gas in the disk is unphysical, making the computed gas pressure fictitious.
In fact, estimates based on dimensional analysis (Shakura
\& Sunyaev 1973) suggest that the most luminous regions of accretion disks
around neutron stars and black holes are dominated by radiation pressure.
The equation of state in these conditions therefore depends in an essential
way on both the heating--cooling balance of the gas and on the diffusion of
radiation.  If the local pressure is wrong, so are any phenomena involving
compressibility, for example, magnetosonic waves.  Moreover, as pointed out
by Shakura \& Sunyaev (1976), there are reasons to think that those regions
of the disk in which radiation pressure dominates gas pressure are thermally
unstable.  If so, neglect of thermodynamics and radiation forces means that
a major dynamical effect has been omitted.
Lastly, we are primarily interested in accretion disks because they can be
so luminous, but ignoring radiative cooling means that no connection between
accretion and its observational manifestations can be drawn.   

        It is for all these reasons that we have embarked upon a program
to introduce self-consistent simulations of accretion dynamics including
dissipation and radiation effects.  In Turner et~al. (2003), we explored
the local dynamical consequences of radiation forces on the development
of the MRI.  Turner (2004) published an illustrative calculation of a
vertically-stratified disk segment, but with only partial energy conservation.
Complete energy conservation in this context was achieved for the first time
in the first paper in this series
(Hirose et~al. 2006), where we presented a study of a shearing-box model of
an accretion disk annulus in which the radiation pressure was about
$20\%$ of the gas pressure.   Here we take the next step by examining
an annulus chosen to have the radiation pressure comparable to the gas
pressure.  In this paper, we concentrate on the thermodynamic properties
of this annulus; in a companion paper (Blaes et~al. 2007), we focus on
its surface structure and dynamics in the vicinity of the photosphere.

        The primary reason we chose to look at this intermediate case
before plunging into the fully radiation-dominated regime is the possibility
of thermal instability when the radiation pressure is truly dominant.  Assuming
that the vertically-integrated dissipation rate is directly proportional
to the vertically-integrated total pressure, Shakura \& Sunyaev (1976) predicted
linear instability when the radiation pressure is more than 0.6 of the total
pressure.  It will therefore be very useful to gather data capable of testing
that fundamental assumption about the relation between dissipation rate and
pressure.

        There is also a second reason to study the borderline case first.
In order to determine by numerical simulation
whether any sort of equilibrium is stable or unstable requires construction of
an initial state that is in equilibrium with respect to the relevant physical
processes.   When the issue is dynamical stability, finding such an equilibrium
is usually straightforward, because the forces can often be written as
explicit functions of the state variables and position.  Here, however, thermal
equilibrium is the result of a balance between diffusive radiative cooling
and heating due to dissipation of turbulence.  Turbulent dissipation
is inherently chaotic and cannot therefore be written as
an explicit function of anything.  If the characteristic timescale of its
fluctuations were short compared to the thermal equilibration timescale,
one could obtain a good approximation by averaging over these fluctuations.
For MRI-driven turbulence, though, this is not the case---previous
work (e.g., Hirose et~al. 2006) has shown that order unity fluctuations
in the global heating rate can persist for as long as 1--2 cooling times.
Thus, it may be intrinsically impossible to select an initial state whose
thermal properties, absent instability, would remain even approximately
steady for several cooling times.

        By studying a disk at the edge of what might turn out to be a
thermally unstable state, we hope
to avoid some of these problems.   A disk segment that goes through large
amplitude fluctuations without exponential runaway will allow us to
explore scaling relations between thermal properties and other physical
parameters.  We hope, too, that the range of states exhibited in these
conditions will provide guidance for choosing an initial state (or states)
for future simulations of genuinely radiation-dominated shearing boxes.

\section{Calculational Method}

      Our basic tool is the code described in Hirose et~al. (2006), although
we have made certain technical changes to improve efficiency.  This code
solves the equations of 3-d Newtonian MHD in a shearing box.  In order to
detect when grid-scale dissipation occurs and capture it locally into heat,
equations for both internal energy and total energy are solved in parallel.
Any discrepancy detected when comparing these two equations we define as
``dissipation".  Because the dependent variable updates are done in an
operator-split fashion, we can distinguish several subcategories.
Energy differences found during the step when both the magnetic field is
updated via the Faraday equation and the velocity is updated for response
to $\vec J \times \vec B$ force we ascribe to ``numerical resistivity";
similar discrepancies discovered during the velocity updates associated
with gas pressure gradients and advection we call
``numerical shear viscosity"; ``numerical bulk viscosity" is the dissipation
associated with the artificial viscosity necessary for a proper treatment
of shocks.  Note that because the fluid acceleration due to magnetic forces
is considered together with the change in the field itself, the
$\vec J \cdot \vec E$ term in Poynting's Theorem automatically identically
cancels.  Both ``numerical resistivity" and ``numerical shear viscosity"
arise from discretization error and round-off error, not from any explicitly
physical effect.  On the other hand, we expect them to be correlated
in location and time with physical dissipation because numerical errors tend
to be largest where the gradients are least well-resolved, and true kinetic
dissipation is often associated with sharp gradients.
Radiation is coupled to the gas via both Thomson scattering and free-free
opacity, the latter treated in a Rosseland-mean sense.  We solve the
radiation transfer problem in the gray flux-limited diffusion approximation.
Because we treat radiation transfer directly (albeit approximately), the
equations we solve incorporate dissipative effects associated
with photon diffusion such as Silk damping.

      The parameters characterizing the problem are the rotation rate at
the center of the box $\Omega$ and the surface density $\Sigma$.  We choose
them on the basis of a Shakura-Sunyaev disk model with central black hole
mass $6.62 M_{\odot}$, a guessed accretion rate that would yield a luminosity
0.1 of the Eddington rate if the efficiency were 0.1, a radius
$r = 150r_g$, and a guessed Shakura-Sunyaev stress ratio $\alpha = 0.02$.
For these parameters, $r = 1.5 \times 10^8$~cm and $\Omega = 17$~s$^{-1}$.
The effective temperature of such a disk segment is $9.0 \times 10^5$~K.
We choose a surface density of $4.7 \times 10^4$~gm~cm$^{-2}$, the value predicted
by a Shakura-Sunyaev model in the gas pressure-dominated limit.
Our unit of length $H=3.1 \times 10^6$~cm is the (half)-disk thickness
in the gas pressure-dominated limit, as given in Krolik (1999).
The midplane optical depth
is $(1.58 \pm 0.01)\times 10^4$, varying slightly due to the
temperature-dependence of the free-free contribution.  At all times,
the opacity is dominated by Thomson scattering.

       Our initial condition is in approximate hydrostatic and radiative
equilibrium, based upon the assumption that the dissipation is proportional
to gas density, but normalized to the total flux emitted by the disk.  To
construct the initial state, we first solved for the equilibrium below
the photosphere,
describing the radiation density by the diffusion approximation.  We
then matched this solution to one above the photosphere in which the
flux is held constant, the density is fixed at the density floor
($10^{-5}$ the initial midplane density), and the gas temperature is
set at the effective temperature.  Even though this region is optically
thin, in our initial condition we let the radiation energy density
decline linearly with height according to $dE/dz = -3 \chi \rho F / c$.

The magnetic field in the initial state is a twisted azimuthal flux tube
with uniform interior magnetic field strength, similar to the initial
magnetic field geometry used in Hirose et al. (2006).  The radius of the
flux tube is $9H/32$.  The maximum poloidal component of the field is one
quarter the magnitude of the total magnetic field, and the initial midplane
plasma $\beta$ (the ratio of gas plus radiation pressure to magnetic energy
density) is 25.

       The computational box extends $\pm 6H$ from the midplane, has thickness
$(3/4)H$ in radius, and $3H$ in the azimuthal direction.  We chose this height
for the box so that the disk photosphere would nearly always be found within
the problem volume, but we would not waste too many cells studying regions
outside the photosphere.  This box is divided
into 32 cells in the radial direction, 64 in
the azimuthal, and 512 in the vertical direction, so that $\Delta r = 0.0234H$,
$r\Delta \phi = 0.0469H$, and $\Delta z = 0.0234H$.

       Boundary conditions at the azimuthal extremes of the problem
area are purely periodic.  The inner and outer
radial surfaces are matched by shearing periodic boundary conditions
(Hawley et~al. 1995).  At the top and bottom surfaces, the boundary
conditions are outflow extrapolations.  As described in Hirose et~al.
(2006), smooth extrapolation of the electric field followed by use of
the induction equation to find the magnetic field ensures that
$\nabla \cdot \vec B = 0$ even in the ghost cells, but permits discontinuities
in $\vec B$.  In order to curtail rare (roughly once every 50--100
orbits) magnetic field discontinuities strong enough to cause a code crash,
Hirose et al. found it necessary to add a small amount of artificial
resistivity to the ghost cells; in this simulation, we found it necessary
to extend this artificial resistivity into the cells in the problem volume
near the top and bottom boundaries.  As for any diffusivity, the natural
unit for this resistivity is the square of the grid-scale divided by the time-step;
to keep the resistivity small, we set its maximum value (achieved only
at the top and bottom surfaces of the box) to
$$\eta_0 = 0.003\left[\min(\Delta x^2,\Delta y^2, \Delta z^2)/\Delta t^2\right].$$
The resistivity is then tapered into the box according to:
$$\eta = (\eta_0/2)
          \left\{\sin\left[(\pi/2)\left(16|z|/L_z - 7\right)\right] + 1\right\}
          \cases{1 & $|z - L_z/2| < L_z/8$ \cr
                 1 & $|z + L_z/2| < L_z/8$ \cr
                 0 & otherwise\cr}.$$
To test whether this artificial resistivity introduces any noticeable
artifacts into the results, we ran a test simulation with a similar scheme
extending only half as many cells from the boundary.  Starting from our
primary simulation's state at $t=55$~orbits,
this test ran for another 55~orbits.  Because of the chaotic nature of MHD
turbulence, there is no reason to expect the test simulation to reproduce
exactly the original one.  However, they are statistically very similar.
Their time-averaged density and pressure profiles, for example, are nearly
identical, and the character of their volume-integrated dissipation and
radiation flux fluctuations are likewise qualitatively consistent.

       Roughly 10 orbits were required for the MHD turbulence to saturate and to
eliminate initial transients.  The simulation ended after 318 orbits.  As
we will show in a later section, this duration amounted to $\simeq 43$ cooling
times.

\section{Results}

\subsection{Global measures of the energy budget}

\begin{figure}
\epsscale{1.0}
\centerline{\psfig{file=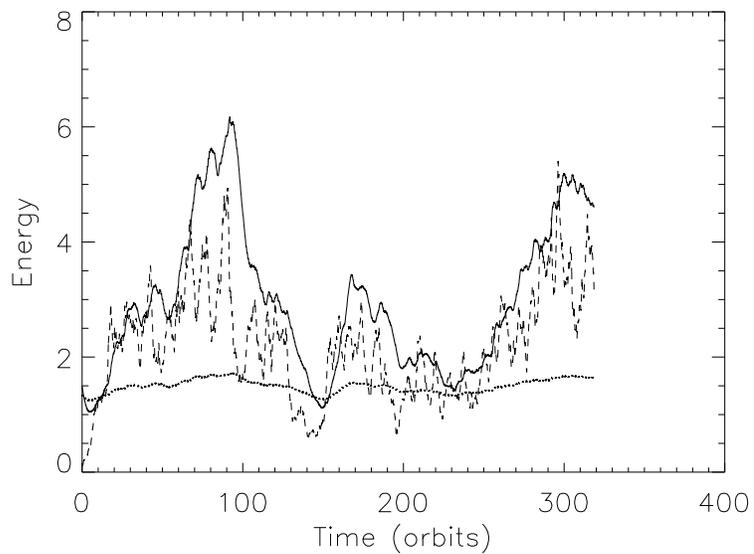,angle=90,width=4.5in}}
\caption{Volume-integrated energy in radiation
(solid curve), gas internal energy (dotted curve), and magnetic
energy (dashed curve; actual value multiplied by 10).
The units of energy are $10^{20}$~erg~cm$^{-2}$.  We do not plot
the kinetic energy because it is much smaller---typically 1--$2\%$
of the total energy.
\label{fig:energyhistory}}
\end{figure}

        The easiest way to obtain an overview of the simulation's
behavior is to look at the history of the volume-integrated energy stored
in gas, radiation, and magnetic field (Fig.~\ref{fig:energyhistory}).
As this figure shows, although the gas internal energy varies rather little
(the maximum is only $1.4\times$ the minimum), the radiation and magnetic
energies vary much more: their peak/trough ratios are 5.9 and 4.8, respectively.
Compared to the initial energy content of the box, the time-average is
$57\%$ greater, but the large fluctuations make that an ill-defined number.
Typical durations of radiation energy fluctuations are $\sim 40$--50~orbits;
although the magnetic energy fluctuations have some power on timescales as
long as this, they also have significant variations over times as short
as $\simeq 10$~orbits.   All the energies are closely correlated: the
cross-correlation between the radiation energy and the gas internal energy
peaks at essentially zero lag; the magnetic energy leads the radiation
energy by $\simeq 5$~orbits.  Both of these facts are easily understood:
the gas and the radiation are very closely coupled, equilibrating in
temperature throughout most of the disk very rapidly.  On the other hand,
increases in radiation energy follow very closely upon increases in the
magnetic field energy, as dissipation of magnetic field provides the
majority of the energy available for radiation.

     Despite these large amplitude variations and the fact that
the radiation energy can be as much as four times the gas energy for as
long as several tens of orbits, there is no
long-term trend or global runaway.  Rather, the system appears to go through
irregular limit-cycle oscillations.  There are two episodes, each 30--40
orbits long, when the radiation pressure accounts for 0.6--0.7 of the
total pressure.  Contrary to the prediction of Shakura \& Sunyaev (1976)
that a radiation pressure fraction of 0.6 is the threshold for instability,
what follows both peaks in radiation/gas pressure ratio is a return to
lower energy content.

\begin{figure}
\epsscale{1.0}
\centerline{\psfig{file=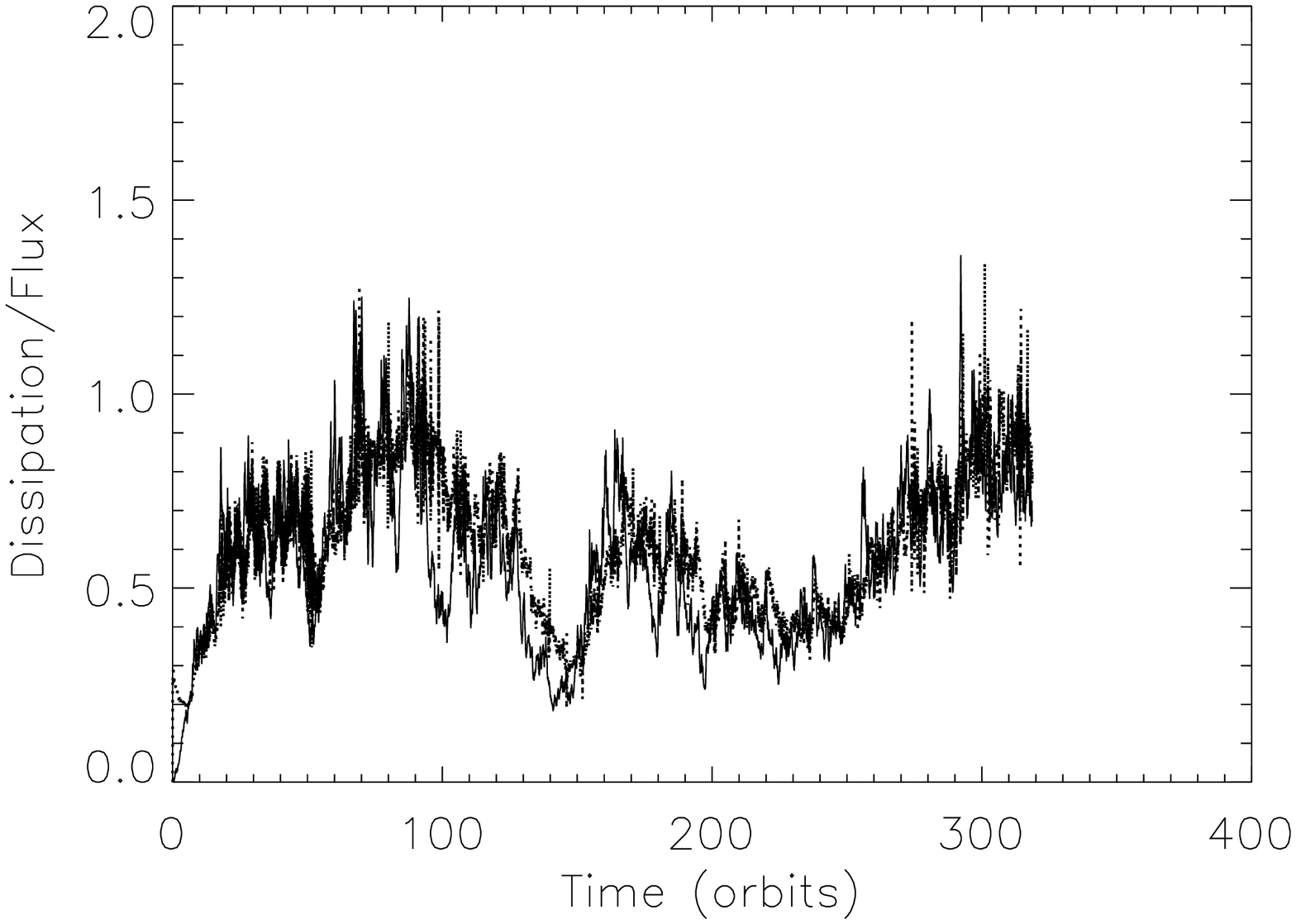,angle=90,width=4.5in}}
\caption{Volume-integrated dissipation rate (solid curve)
and total radiation flux leaving the volume (dashed curve).
Both are in units of $10^{20}$~erg~cm$^{-2}$~orbit$^{-1}$.
\label{fig:fluxdisshistory}}
\end{figure}

      A similar story is told by the history of the dissipation rate and
radiation flux (Fig.~\ref{fig:fluxdisshistory}).  On timescales of $\sim 10$
orbits or more, these two are nearly indistinguishable.  This is, of course,
because the thermal equilibration time $t_{\rm cool}$ is roughly this long.
More precisely, if we define $t_{\rm cool} \equiv (e + E)/F_r$,
where $e$ and $E$ are the gas and radiation energies, respectively, and
$F_r$ is the radiation flux, its mean value is 7.4~orbits, and it varies
only within a rather restricted range: after the initial transients decay,
the instantaneous value of $t_{\rm cool}\simeq 5$--10~orbits except for
a couple of dozen times out of the $\simeq 32,000$ time-steps whose data
we recorded; even then, the maximum excursion is only to $\simeq 13$~orbits.
The variations within this factor of two range appear almost completely
uncorrelated with the state of the disk as defined, for example, by its
total energy.  Thus, the periods during which the radiation pressure
satisfied the nominal thermal instability criterion lasted for $\simeq 4$--5
cooling times, roughly one $e$-folding time for the putative instability.

      In the famous $\alpha$ {\it ansatz} of Shakura \& Sunyaev (1973), both
stress and dissipation are assumed to be proportional to total pressure when all
quantities are vertically-integrated.  We find that the stress in
our simulated volume varies on the thermal timescale and in rough terms
increases with increasing total pressure.  However, there are many order
unity departures from simple instantaneous proportionality, so that the
range of variation is somewhat greater: the maximum is roughly
six times as great as the minimum.  The radiated flux varies in
closely-corresponding fashion, producing a time-averaged
flux that is $65\%$ greater than that associated with the initial condition.
Phrasing these results in $\alpha$ language, we find that the nominal
time-averaged $\alpha$ parameter is $\simeq 0.03$, but its instantaneous
value ranges from $\simeq 0.015$--0.055.

\begin{figure}
\epsscale{1.0}
\centerline{\psfig{file=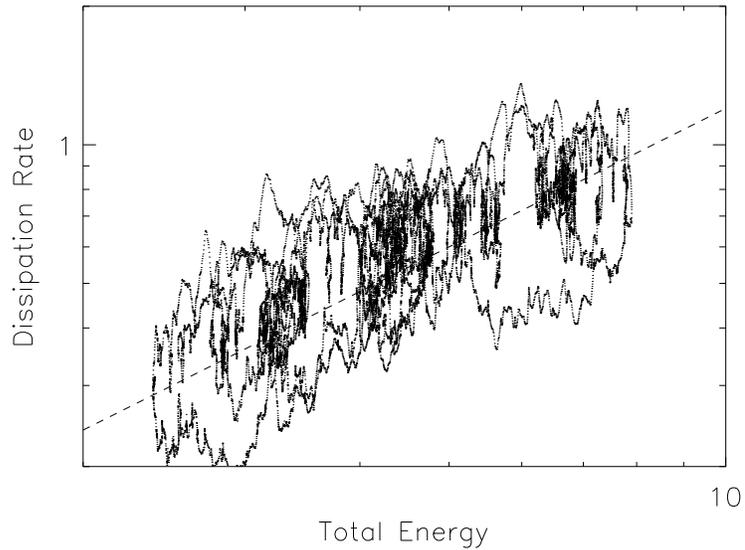,angle=90,width=4.5in}}
\caption{Vertically-integrated and horizontally-averaged dissipation rate versus
vertically-integrated and horizontally-averaged total energy, sampled 100 times
per orbit.  Both are in units of $1 \times 10^{20}$~erg~cm$^{-2}$~orbit$^{-1}$.
The dashed line shows a line of slope unity. \label{fig:energydiss}}
\end{figure}

      As illustrated in Figure~\ref{fig:energydiss}, our simulation
roughly---but only roughly---reproduces the $\alpha$-model expectation
that the dissipation rate should be proportional to total pressure at all times.
Discussing this relationship in terms of total energy rather than total pressure
is advantageous because it slightly simplifies a point we are about to make.
In the limit of either complete gas or radiation dominance, the distinction
would make no difference; here, where the two are comparable, the total
pressure and total energy are almost, but not quite proportional to one
another: $p_{\rm tot} = E/3 + 2e/3$.

Over the entire span of energies and dissipation rates sampled, the single
best-fit power-law relating the two is close to linear.  However, there is
also clear curvature, in the sense that the slope of the dissipation rate
(or stress) with respect to the total energy becomes progressively shallower
with increasing total energy.  In addition, despite the overall correlation,
there are many order unity (factor of 2) departures from the mean trend line.
Placing the data from our earlier simulation (the one published in Hirose
et~al. 2006) on this plot, we find that its relatively small dynamic range
in total energy occupies roughly the bottom $10\%$ in logarithmic range
shown in these data, while its dissipation rate largely overlaps the range
displayed by these new data at that total energy, albeit displaced to slightly
lower values.

\subsection{Internal Structure\label{sec:internal}}

    The internal profiles of density and the various pressures (gas, radiation,
magnetic) vary significantly
depending on the total energy content of the disk.  In order to illustrate
how large the contrast is, we divide the data into two groups: those times
when $e + E > \left[\max(e+E)\right]^{0.75} \left[\min(e+E)\right]^{0.25}$,
which we designate ``high total energy" and those times when
$e + E < \left[\max(e+E)\right]^{0.25}\left[\min(e+E)\right]^{0.75}$,
which we designate ``low total energy".  In the following, whenever we show
results averaged over ``times of high total energy" and ``times of low total energy",
we use this definition.  For maximum contrast between instantaneous properties,
we will use $t=90$~orbits, the time of highest total energy and $t=150$~orbits,
the time of least total energy.

\begin{figure}
\epsscale{1.0}
\centerline{\psfig{file=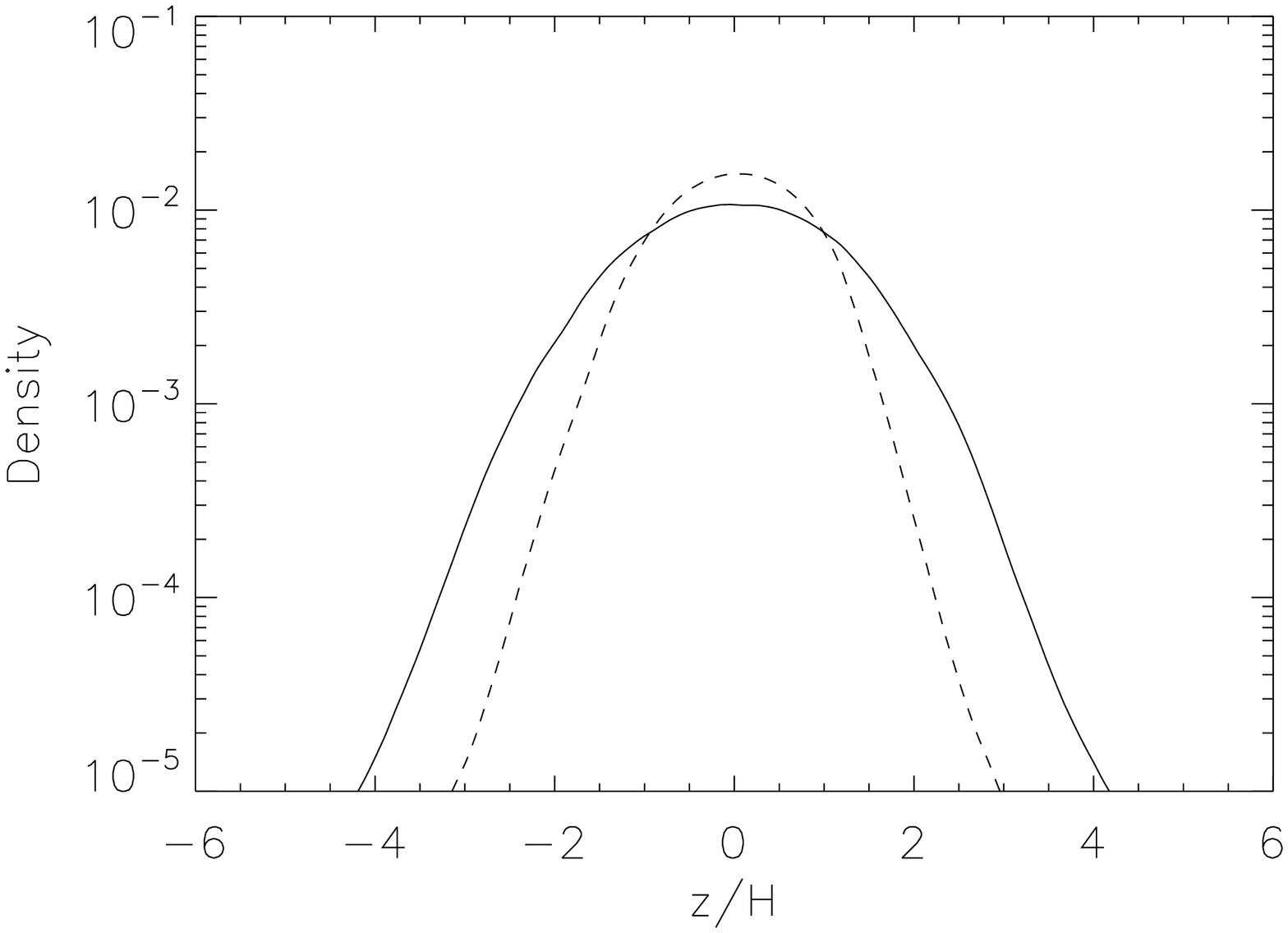,angle=90,width=4.5in}}
\caption{Mean density profile when the total energy is high (solid curve)
and low (dashed curve).  Units of density are g~cm$^{-3}$.
\label{fig:densprofiles}}
\end{figure}

     We begin by showing the density profile (Fig.~\ref{fig:densprofiles}).
At times of high energy content, the disk's matter can extend to higher altitude
away from the midplane.  As a result, its central density falls by $\simeq 30\%$
while the density scaleheight ($h \equiv \int dz \rho(z)|z|/\int dz \rho(z)$)
increases by nearly $50\%$, from $0.6H$ to $0.88H$.

\begin{figure}
\epsscale{1.0}
\centerline{\psfig{file=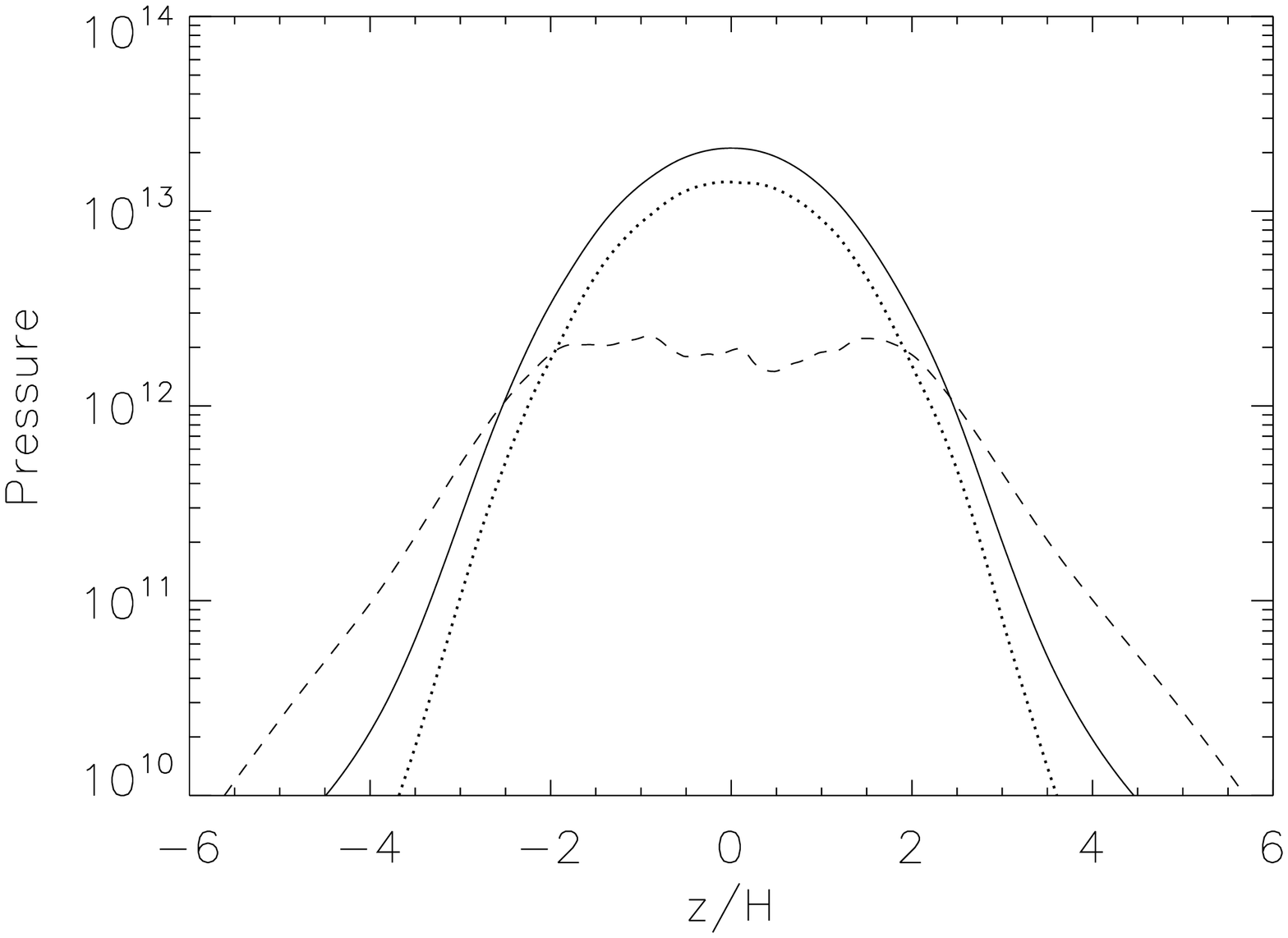,angle=90,width=2.5in}
\psfig{file=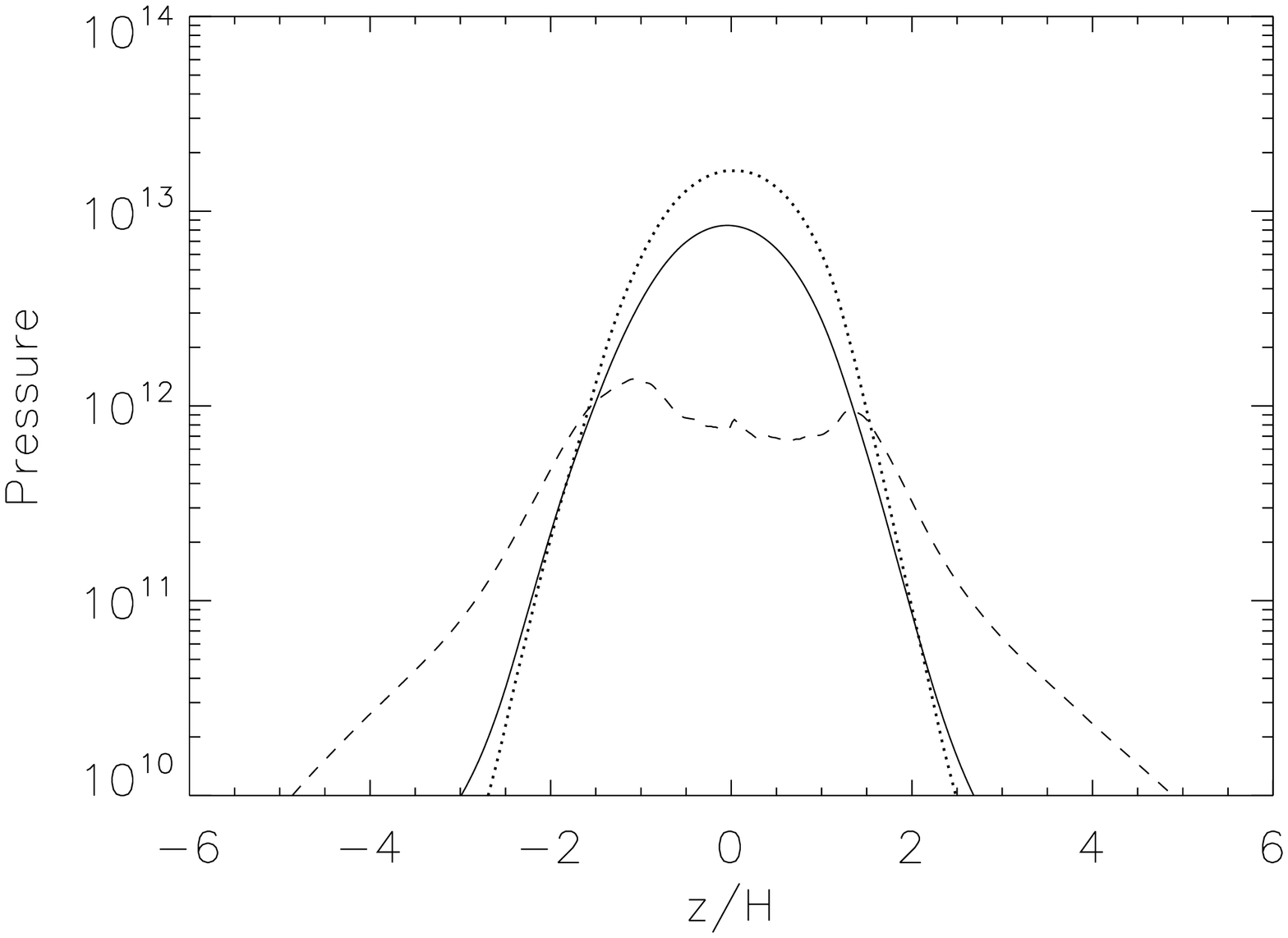,angle=90,width=2.5in}}
\caption{Left panel: Mean radiation pressure profile (solid curve),
gas pressure profile (dotted curve), and magnetic energy profile (dashed curve)
when the total energy is high.  Right panel: the same quantities when the
total energy is low.  Units for all pressures are dyne~cm$^{-2}$.
\label{fig:pressprofiles}}
\end{figure}

     Not surprisingly, the pressure profiles show a similar pattern of
broadening at times of high energy content (Fig.~\ref{fig:pressprofiles}).
At all times, the magnetic pressure is roughly constant across the middle
several scale-heights, while the gas and radiation pressures are centrally
peaked.   Consequently, the pressure becomes magnetically-dominated at high
altitude.  The height of the transition point rises with increasing total
pressure, from $\simeq 1.5$--$2H$ at times of low pressure to $\simeq 2.5$--$3H$
at times of high pressure.  In the midplane, the dominant contributor to the
pressure changes as a function of total energy: at low total energy, the
gas pressure $p_g$ is greater than the radiation pressure $p_r$
in the midplane, whereas it is the other way around at times of high energy content.
This overall pattern, in which a combination of gas and radiation pressure
dominates in the central several scale-heights, while the magnetic pressure
has a profile that is flat in the body of the disk but declines slowly
at higher altitudes so that it dominates the regions well away from the plane,
is very much like what is seen in the purely gas-dominated case (Hirose et~al. 2006).

    This characteristic pattern for the magnetic field strength may be
related to the way the magnitude of the fastest-growing MRI wavelengths
depends on height.  Both for axisymmetric and nonaxisymmetric MRI, the
fastest growing modes have vertical wavelengths $\lambda \sim v_A/\Omega$
(Kim \& Ostriker 2000), provided there is some nonzero poloidal field and
the equilibrium background changes only over longer scales.
We find that both $v_A/\Omega$ and the gas scale-height $c_t/\Omega$ are
very nearly constant across the central disk region within which the magnetic
field varies little with height, with ratio $v_A/c_t \simeq 0.3$.  Here
$c_t$ is the sound speed of acoustic waves supported by both gas and radiation
pressure.  Just below the layer where the magnetic field begins to decline
upward, the ratio $v_A/c_t$ starts to increase sharply with altitude
(as we discuss immediately below, radiation pressure decouples from
acoustic waves as the gas density falls; if our estimate were adjusted
to account for the resulting diminution of $c_t$, this effect would be
enhanced).  This pattern may indicate where the MRI can operate successfully\
and where it cannot.

\begin{figure}
\epsscale{1.0}
\centerline{\psfig{file=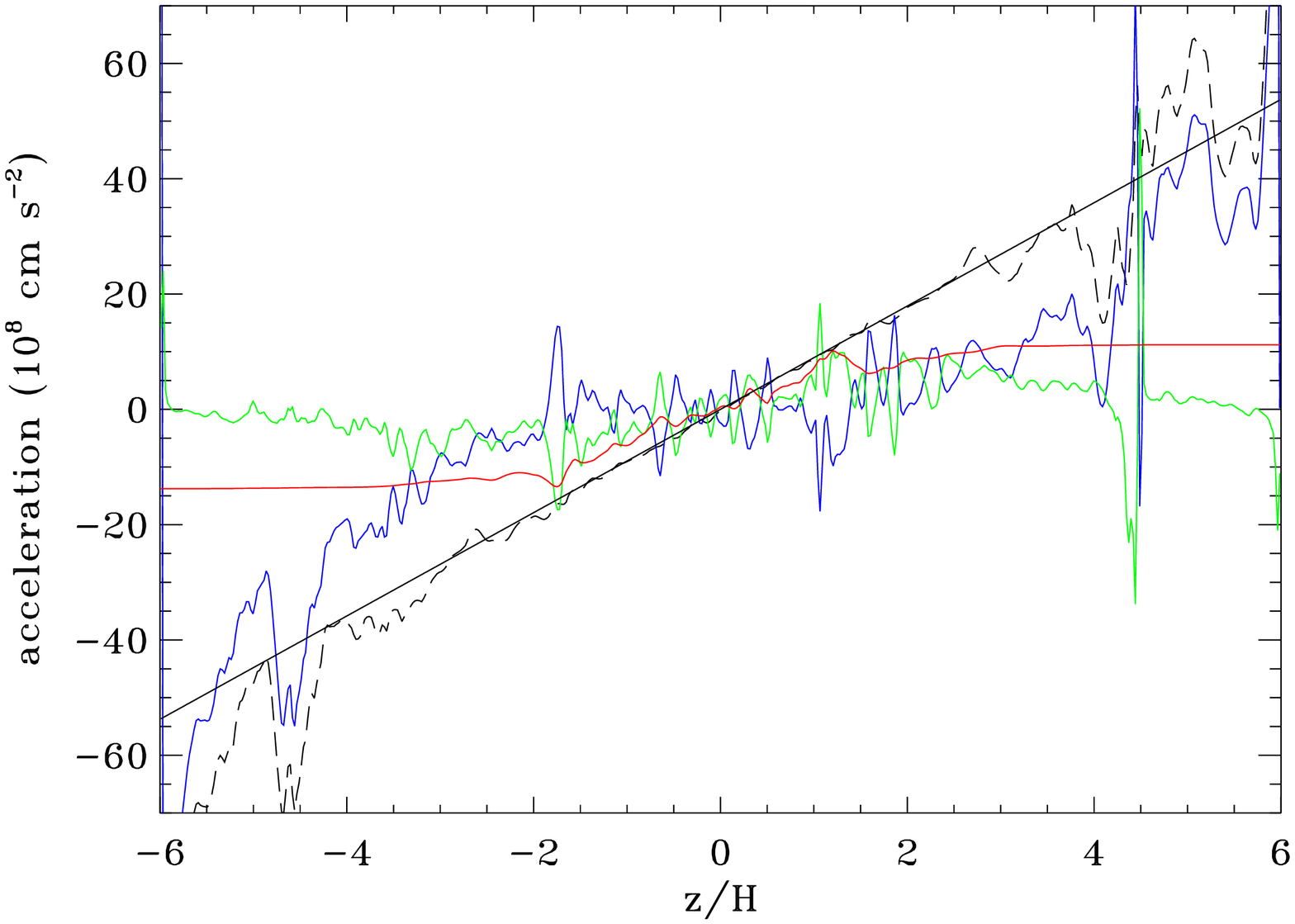,width=2.5in}
\psfig{file=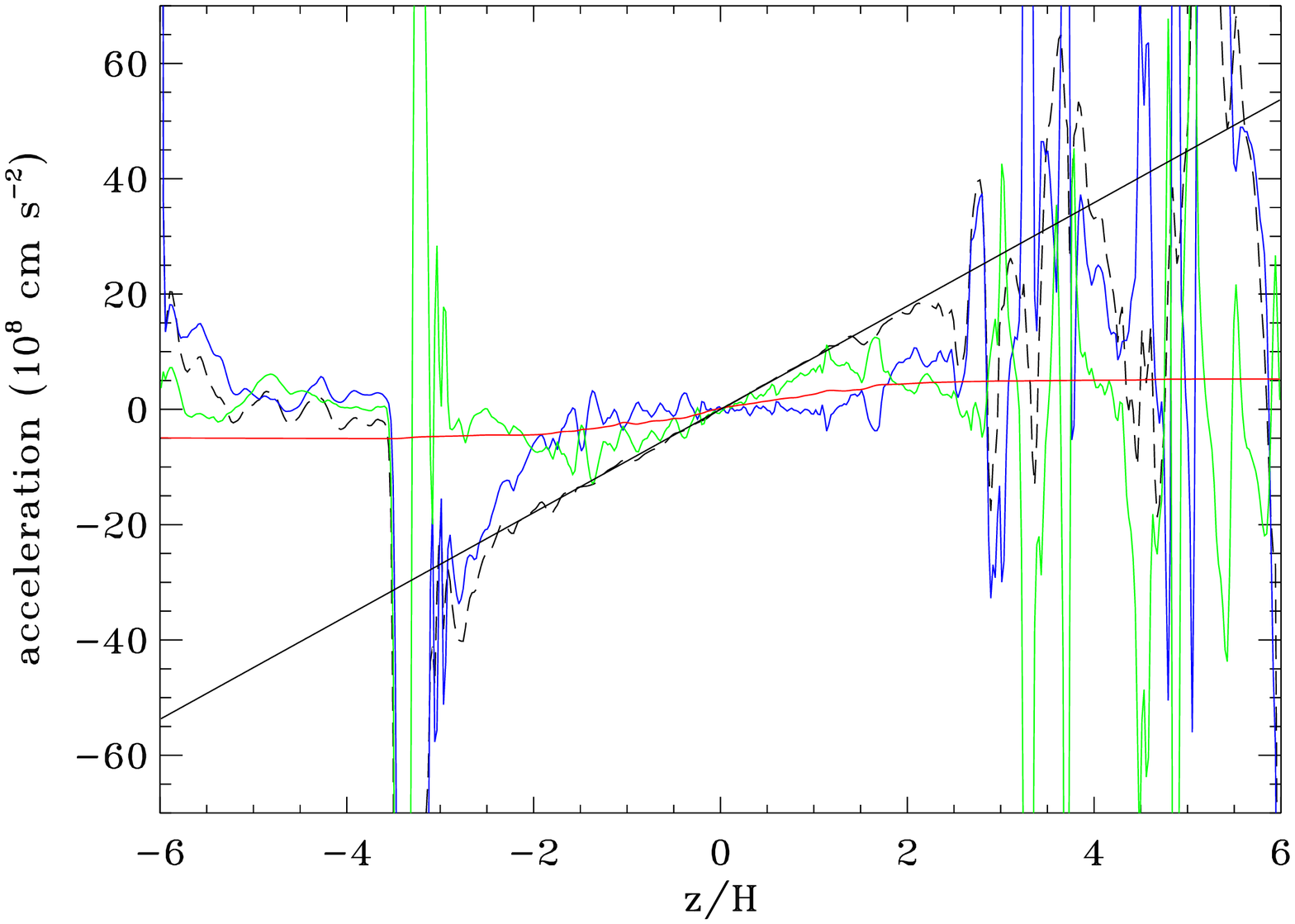,width=2.5in}}
\caption{Horizontally averaged contributions to the outward vertical force
per unit mass acting on the gas at $t=90$ orbits (high total energy, left
panel) and $t=150$ orbits (low total energy, right panel).
The different contributions are from gas pressure gradients
(green), radiative acceleration (red), and magnetic pressure gradients
and tension forces (blue).  The dashed black curve gives the total vertical
acceleration from these forces and the solid black curve shows the inward 
gravitational acceleration.}
\label{fig:vertacc}
\end{figure}

   For a more detailed look at which forces support the gas,
Fig.~\ref{fig:vertacc} shows the volume-weighted horizontally-averaged
vertical force per unit
mass from gas, radiation, and magnetic forces, at the highest and lowest
total energy epochs.  In the former case, hydrostatic balance is approximately
established at all heights in the box, with magnetic forces dominating the
gas and radiation pressure forces outside $3-4H$.  In contrast, the outer
regions are not in hydrostatic balance in the low energy epoch.  At these times,
there is no significant support against gravity for $z\lesssim-4H$, and
strong fluctuations in gas pressure and magnetic forces exist for $z\gtrsim3H$.
As can be seen from Figure~\ref{fig:energyhistory}, the two periods during
which the energy content is particularly low are both relatively short-lived;
the failure to achieve equilibrium is presumably due to the brevity of
these episodes.  During those times, matter that had formerly been
at high altitude falls rapidly toward the midplane once radiation support
of material closer to the midplane has been removed.  Strong shocks
form at intermediate altitudes where this gas strikes denser material.


\begin{figure}
\epsscale{1.0}
\centerline{\psfig{file=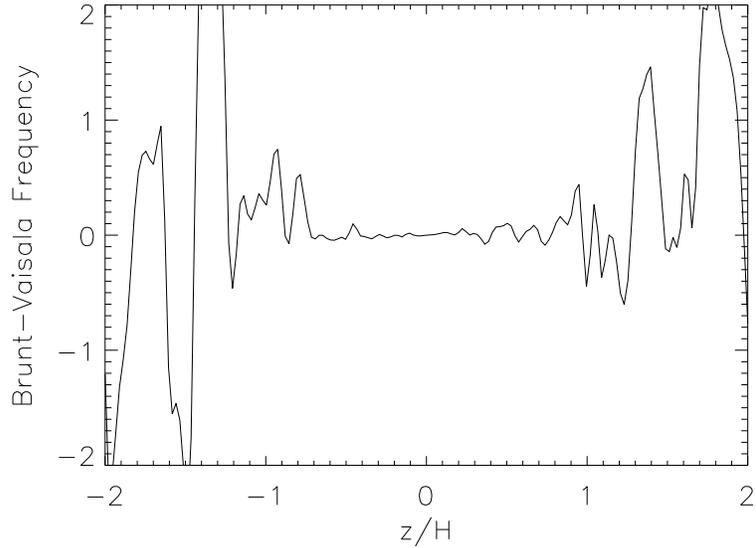,width=4.5in,angle=90}}
\caption{The long-wavelength Brunt-V\"ais\"al\"a frequency in units
of $\Omega^2$,
computed on the basis of horizontally-averaged data at a time
of intermediate energy content, $t=175$~orbits.}
\label{fig:b-v}
\end{figure}

     In the central body of the disk, where magnetic pressure is less
important than either gas or radiation pressure for support against gravity,
dynamical perturbations can be understood primarily in terms of conventional fluid
mechanics (as Fig.~\ref{fig:vertacc} shows, magnetic forces are the primary
support against gravity where $|z| \gtrsim 3H$).  The only special complication
is one due to the importance of radiation pressure: fluctuations behave
differently according to whether photons can diffuse across a wavelength
in a time long or short compared to a dynamical time.  That the cooling
time is $\simeq 7$~orbits immediately implies that the diffusion time
across a scale-length near the disk midplane is $\sim 10$~orbits.  Because
the characteristic timescale for buoyant motions is of order the orbital
period (or rather longer, as shown by Fig.~\ref{fig:b-v}),
long wavelength fluctuations of this variety near the midplane are in the
long-diffusion time regime, while shorter wavelengths near the
midplane and modes with all wavelengths at greater heights are in the
short-diffusion time limit.

      When the wavelength is long enough that the
diffusion time is greater than the dynamical time, gas and radiation are
closely coupled, and the effective Brunt-V\"ais\"al\"a frequency may
be written as
\begin{equation}
N^2 = g(z)\left[\frac{1}{\rho c_{t}^2}
    \frac{\partial}{\partial z}
    \left(p_g + p_r\right) - \frac{\partial\ln\rho}{\partial z}\right],
\end{equation}
where $g(z) = z\Omega^2$ is the vertical component of genuine gravity,
$c_t^2 \equiv \Gamma_1 (p_g + p_r)/\rho$, and $\Gamma_1$ is
Chandrasekhar's generalized adiabatic constant,
\begin{equation}
\Gamma_1 = \frac{16 + 40e/E + 10(e/E)^2}{3(4 + e/E)(1 + 2e/E)}.
\end{equation}
On the other hand, radiation can quickly diffuse out of short
wavelength fluctuations, eliminating any fluctuation in the radiation
pressure on such small scales.   When this is the case, buoyancy
is strongly suppressed, partly because the upward radiation force
partially cancels gravity, and partly because rapid thermal equilibration
between gas and radiation enforces a single density for all gas parcels
at a given height, independent of their initial specific entropy
(Blaes \& Socrates 2003).  At the order of magnitude level, the
growth rate of short wavelength buoyant modes is
$\sim 0.03\tau_H (v_{\rm orb}/c)(H/r) (kH)^{-2}\Omega$, where
$\tau_H$ is the optical depth across a length $H$.  In this
simulation, $(v_{\rm orb}/c)(H/r) \sim 10^{-3}$,
so we expect such short wavelength
fluctuations to play at most a minor dynamical role where the
magnetic pressure is smaller than gas or radiation pressure.

  Evaluating the long-wavelength $N^2$ in terms of horizontally-averaged
quantities at individual times (as shown in Fig.~\ref{fig:b-v}), we
find that $N^2 \sim 0.1\Omega^2$
near the midplane ($|z| \lesssim H$) at almost all times (around
$t=90$~orbits, the time of greatest energy content, $|N^2|$ is slightly
larger).  In other words, in the deepest part of the disk, long
wavelength dynamical perturbations are almost neutrally stable because
the specific entropy is almost constant with height.  At somewhat
greater heights, but still within the central disk body as just defined,
fluctuations with wavelengths long enough to be very optically thick can
have larger $|N^2|$ and may have either sign.  However, because the density
falls with height, the long wavelength case applies to a smaller fraction
of the range of possible wavelengths for $1 \lesssim |z|/H \lesssim 2$
than for $|z|/H \lesssim 1$.   In this region, therefore, nearly all modes
have very low frequencies, whether oscillatory, growing, or damped.
Thus, through most of the disk body, where radiation and gas pressure combine
to outweigh magnetic pressure in support against gravity, only occasionally
do dynamical perturbations grow: for the most part, they are either truly
or neutrally stable.

Our earlier simulation, in which radiation pressure was only $\simeq 0.2$
times as great as the gas pressure, provides an interesting contrast:
in those conditions, where radiation pressure is a minor effect, $N^2$
is likewise very close to zero in the inner few scale-heights, but
rises smoothly toward $\sim 1$ at greater altitude.  Thus, when
gas pressure dominates, convectively unstable zones are still rarer
than they are when gas and radiation pressure are comparable.

Because the gas and radiation pressure contributions lead to almost
neutral buoyancy, magnetic forces are all that is left to drive fluctuating
vertical motions in the disk body, particularly at altitudes
1--$3H$ from the midplane.  As has often been seen in previous
simulations of stratified shearing boxes \cite{ms00,hks06}, regions
of locally strong magnetic field are generated a few scale-heights
from the plane and rise another few scale-heights at a speed $\sim 0.2\pi H\Omega$.
Although this rise temporarily depletes the field at its altitude of
origin, the field intensity is generally restored in 1--2 orbits,
only to generate another rising current.

In the upper layers of the disk, where magnetic pressure predominates,
the situation changes.  There unstable Parker modes can become important.
We discuss the dynamics of these surface layers in the companion paper.

\subsection{Dissipation distributions}

\begin{figure}
\epsscale{1.0}
\centerline{\psfig{file=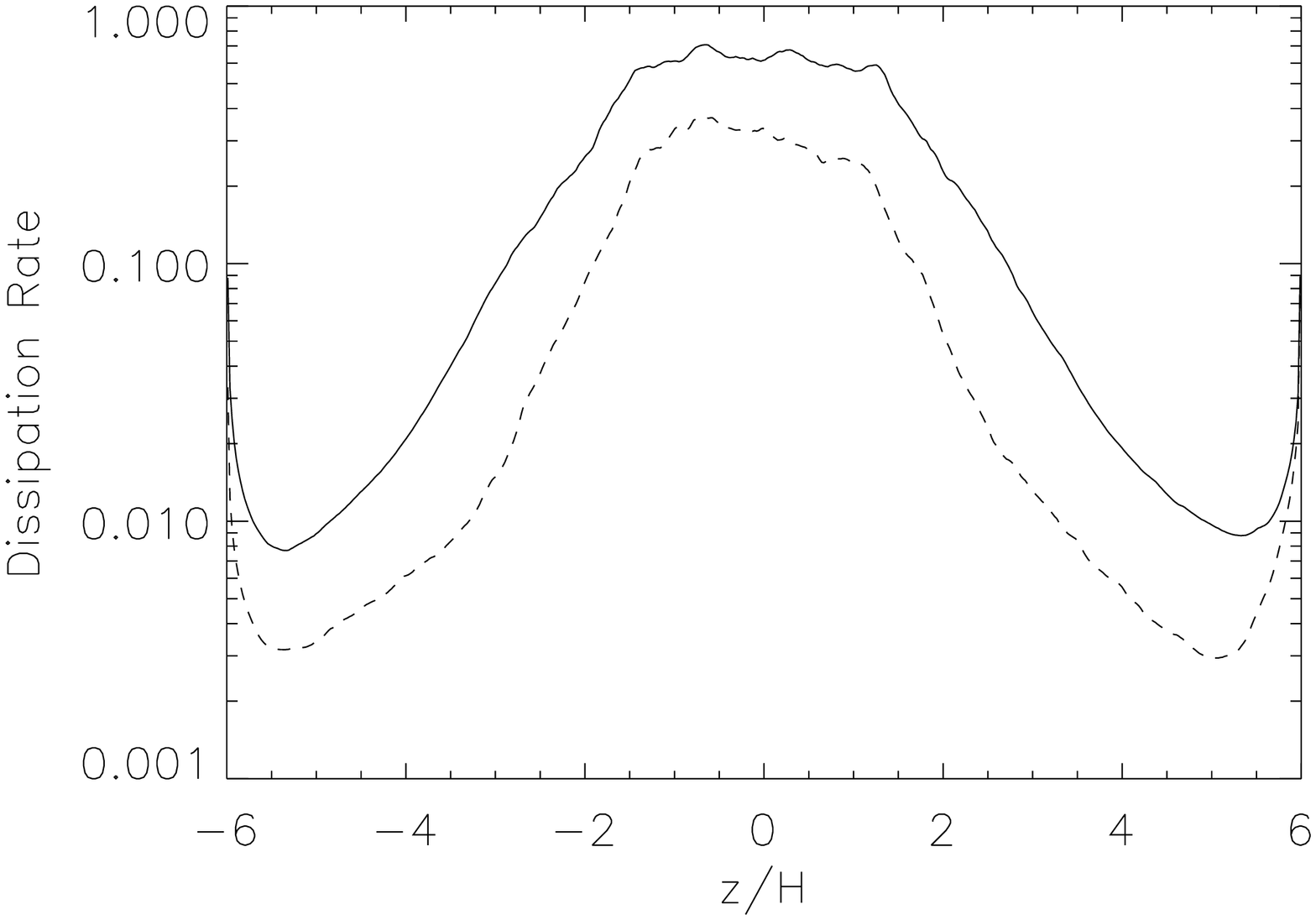,angle=90,width=4.5in}}
\caption{Vertical profile of horizontally- and time-averaged dissipation
per unit volume.  Times of high total energy are shown by a solid curve,
times of low total energy by a dashed curve.  The units of dissipation rate are
$10^{13}$~erg~cm$^{-3}$~s$^{-1}$.
\label{fig:dissprofz}}
\end{figure}

\begin{figure}
\epsscale{1.0}
\centerline{\psfig{file=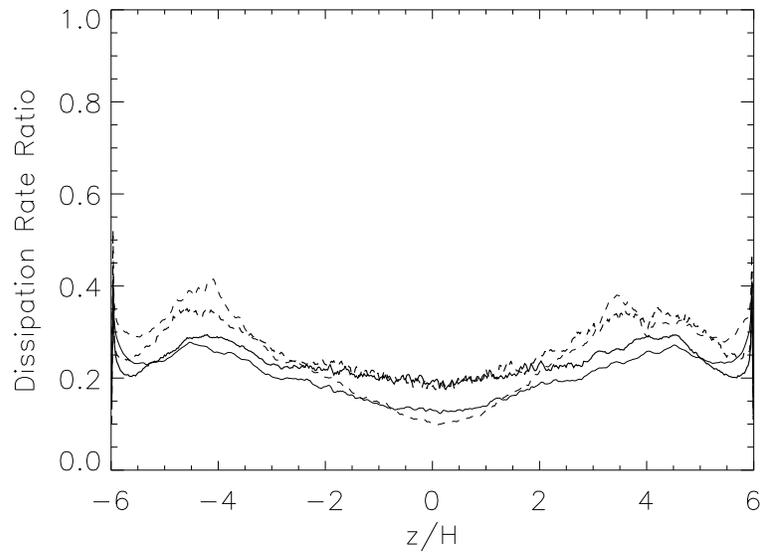,angle=90,width=4.5in}}
\caption{Ratios of the different categories of horizontally- and
time-averaged dissipation rates.  Thick curves are the ratio of
numerical viscosity dissipation to that due to numerical resistivity,
thin curves are the ratio of artificial bulk viscosity dissipation
to that of numerical resistivity.  Solid curves are averaged over
times of high energy content, dashed curves over times of low
energy content.
\label{fig:disrates}}
\end{figure}

    The vertical profile of dissipation per unit volume is also much
like the gas-dominated case, provided
one makes an adjustment for both the higher rate of dissipation that goes along
with conditions of higher pressure and the greater thickness
of the disk.  Figure~\ref{fig:dissprofz} shows this profile, averaged over
the same selection of times used elsewhere in this paper for high total energy
and low.  At all times, the great majority of the dissipation takes place
within 1.5--$2H$ of the midplane, with the width increasing with total
pressure.  Because numerical resistivity is generally (but not always: see
Fig.~\ref{fig:disrates}) the dominant dissipation mechanism,
the characteristic scale for the dissipation profile is mostly a reflection of
the scale of the magnetic field distribution (we will add further detail to
this point below).  Although several thousand sampled times were averaged to create
the high total energy curve, its bumpiness, visible even in a logarithmic plot,
demonstrates the strong intermittency of the dissipation.
The small uptick in dissipation per unit volume just inside the outer boundaries
is likely an artifact of the boundary conditions.

\begin{figure}
\epsscale{1.0}
\centerline{\psfig{file=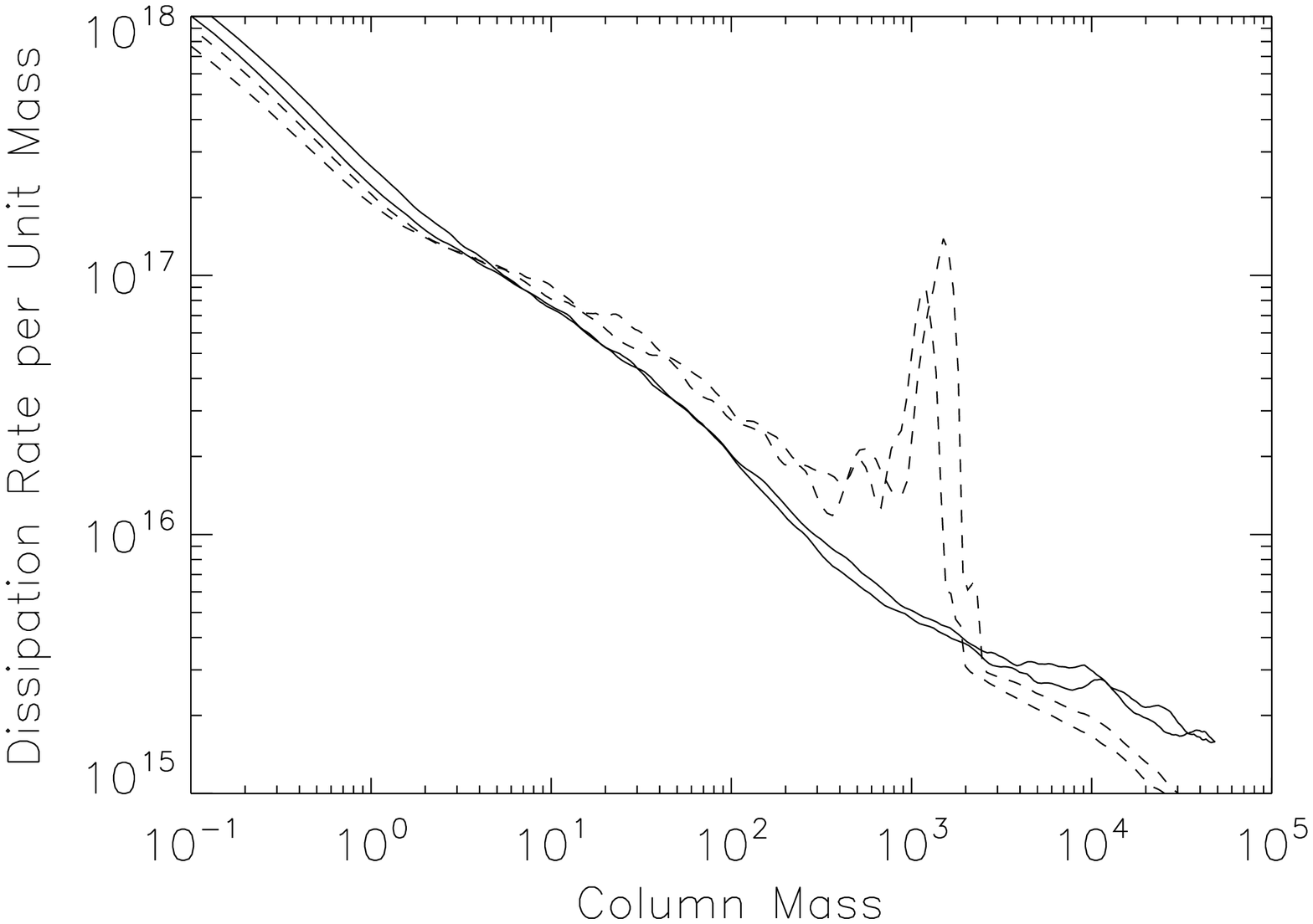,angle=90,width=2.5in}
\psfig{file=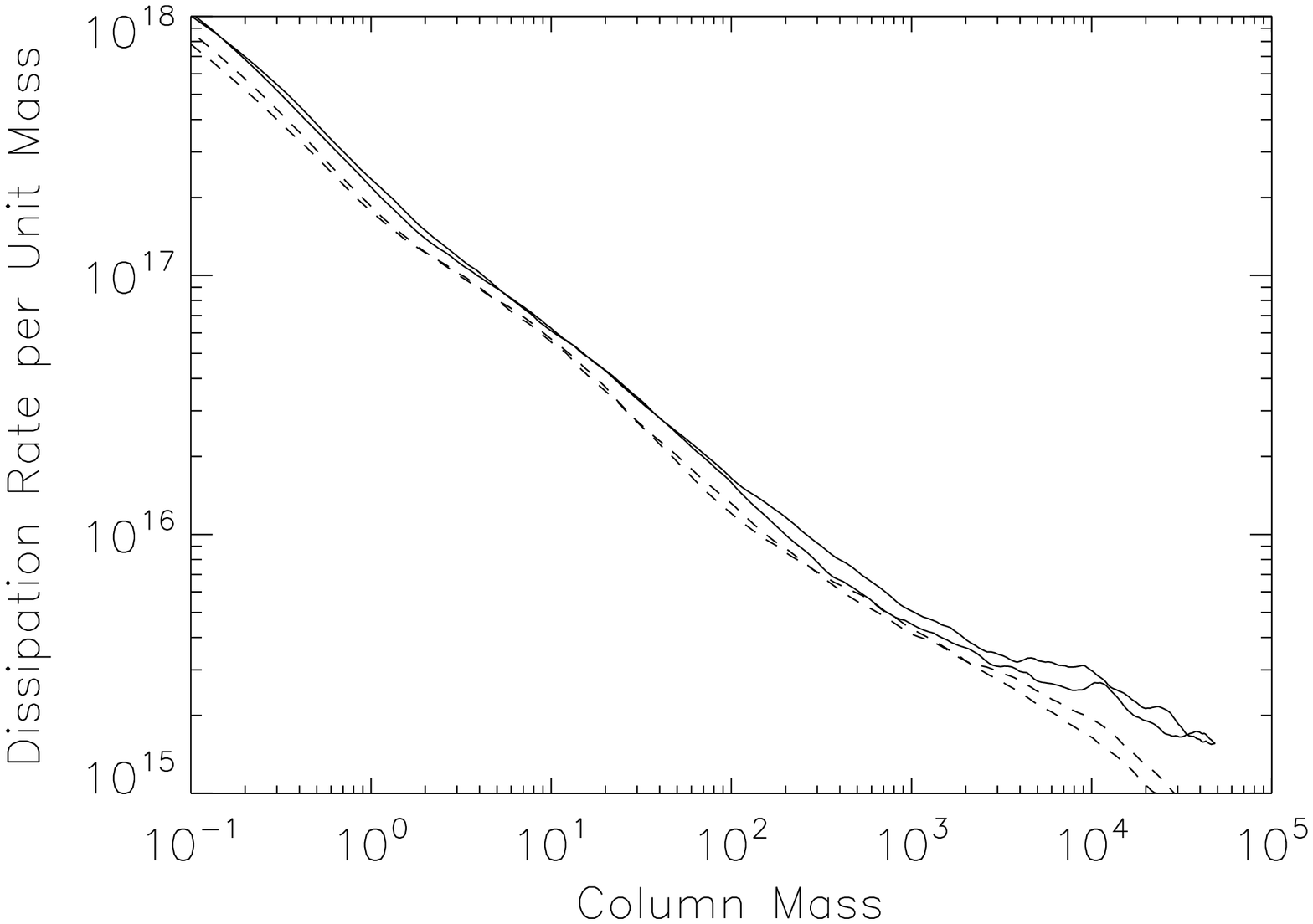,angle=90,width=2.5in}}
\caption{Dissipation per unit mass at times of high total energy (solid curves)
and low total energy (dashed curves) averaged two different ways.  There is
a pair of curves for each case because the top and bottom halves of the disk
are plotted separately.
Left panel: $\langle j(z) \rangle = \langle Q/\rho \rangle$.
Right panel:  $\langle j(z) \rangle = \langle Q\rangle/\langle \rho \rangle$.
Units for specific dissipation are erg~g$^{-1}$~s$^{-1}$, units for column mass
are g~cm$^{-2}$.
\label{fig:disspermass}}
\end{figure}

     Despite the sharp fall with increasing height in the dissipation rate per
unit volume, the dissipation rate per unit mass $j$ consistently
rises outward (Fig.~\ref{fig:disspermass}).  This is a reflection of the fact
that the dissipation profile exhibits a shallower slope than does the gas
density profile.  To a good approximation, $j \propto m^{-1/2}$, where
$m \equiv \int_z^{\infty} dz^\prime <\rho(z^\prime)>$
(for $z < 0$, the limits on the integral are $-\infty$ and $z$, of course).
Note that the gas density used in the column density integral is
horizontally- and time-averaged
in a fashion identical to that of the dissipation, selecting out those times
of either particularly high or particularly low energy content.
Put another way, the dissipation taking place outside a column density $m$
is $\propto m^{1/2}$, so that the total dissipation is dominated by the very
optically thick central part of the disk, but a significant minority takes
place relatively near the surface if distance is measured in mass units.
Interestingly, unlike the
distribution of dissipation per unit volume, for which there is a clear
characteristic scale-height, the distribution of dissipation per unit mass
displays at most only a weak signature of any special column density scale.

The fluctuations in both dissipation and density are great enough
that it matters how one constructs time averages for $j$.  In
Figure~\ref{fig:disspermass},
the left-hand panel shows the time-average of the ratio of instantaneous
horizontal averages of dissipation to density; the right-hand panel shows the
ratio of time-averages.  That is, if the instantaneous horizontal average
of the dissipation per unit volume is $Q(t,z)$ and the instantaneous horizontal
average of the gas density is $\rho(t,z)$, the former quantity is
$\langle Q/\rho \rangle_t$, while the latter quantity is $\langle Q \rangle_t/
\langle \rho \rangle_t$.  Here $\langle \ldots \rangle_t$ denotes a
time-average of the quantity within angle-brackets.   Near
$m \sim 10^3$~g~cm$^{-2}$ (generally $|z| \simeq 2H$, or the lowest region
where magnetic pressure dominates gas and radiation pressure), at
times of low total energy there are a few instances of such a high localized
dissipation rate per unit mass that they color the entire time-average, even
though they do not contribute substantially to the time-average of $Q$ or
$\rho$ separately.  The origin of these episodes of exceptionally high
dissipation per unit mass appears to lie in strong shocks driven by
the vertical collapse of high altitude material when the radiation
content has just been sharply reduced.  Although the mean column
density to the cells in which they take place is $\sim 10^3$~g~cm$^{-2}$,
the actual column density at the time of the shock is often 2--3 orders
of magnitude smaller.

As in our earlier simulation of gas-dominated conditions (Hirose et~al. 2006),
we find that most cells have both very low dissipation rate and very low
current density.  However, the total dissipation is dominated
by the small number of cells with the highest dissipation rate, which tend also
to have the highest current density.  Roughly speaking, $75\%$ of the total
dissipation comes from the cells with current densities above 1/3 the greatest
current density found in the volume; these represent $\sim 10^{-3}$ of the
total number of cells.

The balance between our three different numerical
dissipation mechanisms changes systematically both with total energy
content and with height from the midplane.   Although numerical
resistivity always dominates the horizontal and time averages (ranging
from 60--$80\%$ of the total at all heights, and taking the largest
share in the disk body where the total dissipation rate is highest), the
relative importance of artificial bulk viscosity, as compared to
numerical viscosity, increases by almost a factor of two from
the midplane to $|z/H| \simeq 4$ at times of high energy content
and by roughly a factor of four at times of low energy content
(Fig.~\ref{fig:disrates}).  This shift signals the relative
importance of shocks in the magnetically-dominated zone well away
from the midplane, and also contributes to the trend toward greater
dissipation per unit mass at lower column density to the surface.

\begin{figure}
\epsscale{1.0}
\centerline{\psfig{file=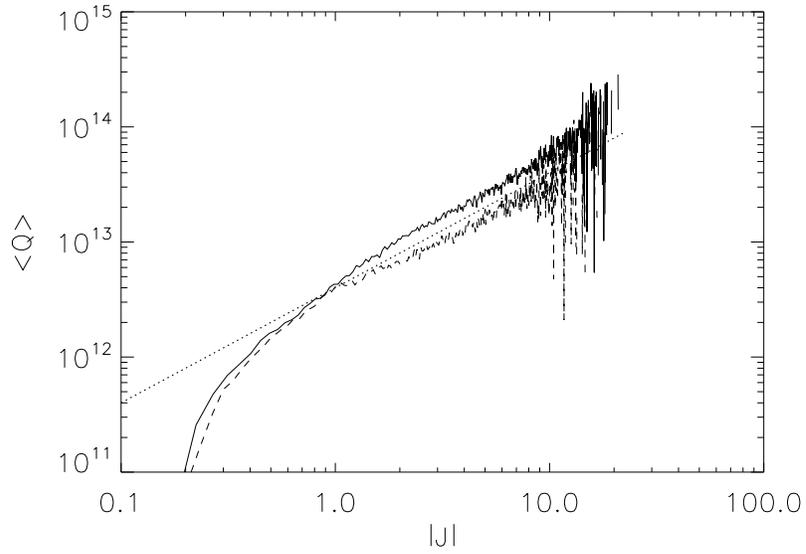,angle=90,width=4.5in}}
\caption{The mean dissipation rate as a function of current density for a moment
of high total energy (orbit 90: solid curve) and a moment of low total
energy (orbit 150: dashed curve).  Current density is shown in scaled units
($H/c$ times cgs); dissipation rate is in erg~cm$^{-3}$~s$^{-1}$.  The
dotted line has slope 1.
\label{fig:disscurrcorr}}
\end{figure}

Averaged over the entire simulation volume, the scaling between current density
and mean dissipation rate in the simulation reported here resembles, but is
slightly different from, the relation found in the gas-dominated case.
We define the mean dissipation rate found at a given current density by
$$\langle Q \rangle \equiv  \int \, dQ \, Q N(Q,|\vec J|)/
                 \int \, dQ \, N(Q,|\vec J|),$$
where $N(Q,|\vec J|)$ is the number of grid cells having dissipation rate
$Q$ and current density $\vec J$.  In the gas-dominated case,
we found $\langle Q \rangle \propto |J|^{1.13}$ at moderate to high current densities
(Fig.~\ref{fig:disscurrcorr}).  Just as before, at small current density
$\langle Q\rangle$ rises steeply with $|J|$ (roughly $\propto |J|^2$), but
at intermediate current levels the relation rolls over so that $\langle Q \rangle$
is approximately $\propto |J|$.
Not surprisingly, given the concentration
of dissipation to the central disk body, the different regions of these
correlations are not equally well-sampled at all altitudes: cells in
the densest part of the disk are much more likely to have high current
densities---and therefore high dissipation rates---than cells at high
altitude.  In addition, cells with low current density near the midplane
tend to have a rather greater dissipation rate than cells with
the same current density nearer the disk surface.

Only subtle changes distinguish the mean current density---dissipation
correlation at low total
energies from those at high.  When the total energy is higher,
the range of current densities found extends to slightly higher values, and
the mean dissipation at a given current density tends to be
somewhat greater, but these contrasts are only at the tens of percent level.
The slope of the dissipation--current density relationship is also slightly
steeper at times of high total energy than at times of low: a linear
least-squares fit to the relation over the upper factor of 50 in current
density finds that the slope rises from $\simeq 1.0$ at low total energy to
$\simeq 1.15$ at high.  The main
factor accounting for the greater dissipation at high energy than at low
is a larger number of cells having high current densities and consequently
high dissipation rates.

Although there is a clear correlation between current density and dissipation
rate, the dissipation rate is by no means a function of current density alone. 
The rms logarithmic scatter around the mean relation corresponds to a factor
of $\simeq 3$--4, declining slightly at the highest current densities.

\begin{figure}
\epsscale{1.0}
\centerline{\psfig{file=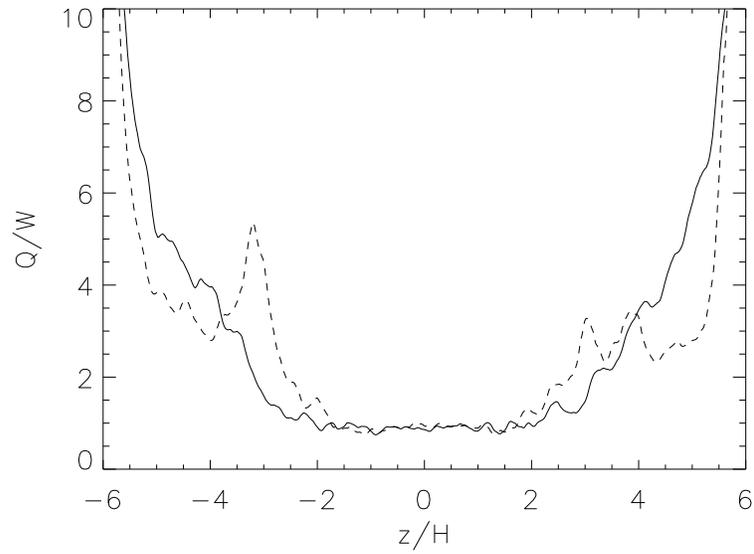,angle=90,width=4.5in}}
\caption{The ratio of horizontally-averaged dissipation rate $Q$ to
horizontally-averaged stress $W$.  An average over times of high
energy content is shown by the solid curve, an average over times
of low energy content is shown with the dashed curve.  The units
are chosen so that $Q/W = 1$ if the stress were produced by a
genuinely dissipative mechanism.
\label{fig:distratio}}
\end{figure}

   It is interesting to contrast these results to those
that might have been expected based on the thought, frequently voiced in
older studies of accretion disk physics, that the stress is produced by some
sort of dissipation process akin to conventional viscosity.  If that were
the case, the local dissipation rate and the local stress should be
proportional to one another with a fixed ratio.  Correlated MHD turbulence,
the mechanism now known to produce the bulk of the stress in accretion
disks, is not immediately dissipative, and therefore does not necessarily have
this property.  Our data strongly confirm the essentially non-dissipative nature
of the stress, as seen in Figure~\ref{fig:distratio}.   Although in the
disk body the horizontally- and time-averaged dissipation and stress are
related approximately in the way predicted for dissipative stress mechanisms
(even after averaging, the ratio of dissipation to stress is about 10--$20\%$
less than this model would predict), their relationship changes strikingly
with increasing distance from the midplane.   Near the disk photosphere,
their mean ratio is a full order of magnitude greater.   The origin of this
shift most likely lies in the increasing importance of shocks in that
magnetically-dominated region (see the earlier discussion in this subsection
and \S~5 in the companion paper).   Moreover, there is little difference
between times of high and low energy content in the way the relative
magnitudes of stress and dissipation change with altitude.  The stress
and dissipation data from the gas pressure-dominated simulation reported
in Hirose et~al. (2006) also show a very similar pattern.

   As a final remark in this section, we caution that the mechanism of
this dissipation is truly numerical, and not physical.  Its association
with current density might best be interpreted as an association with
regions of strong magnetic field gradients.   Where gradients are
strongest, numerical error is greatest when the equations are discretized
on a fixed-size grid.  Our claim to some fidelity to the physics is
based on the fact that many genuine physical dissipation mechanisms
are also stronger in regions where the gradients are large, and on
the very accurate energy conservation the code provides.

\subsection{Vertical Energy Transport}

\begin{figure}
\epsscale{1.0}
\centerline{\psfig{file=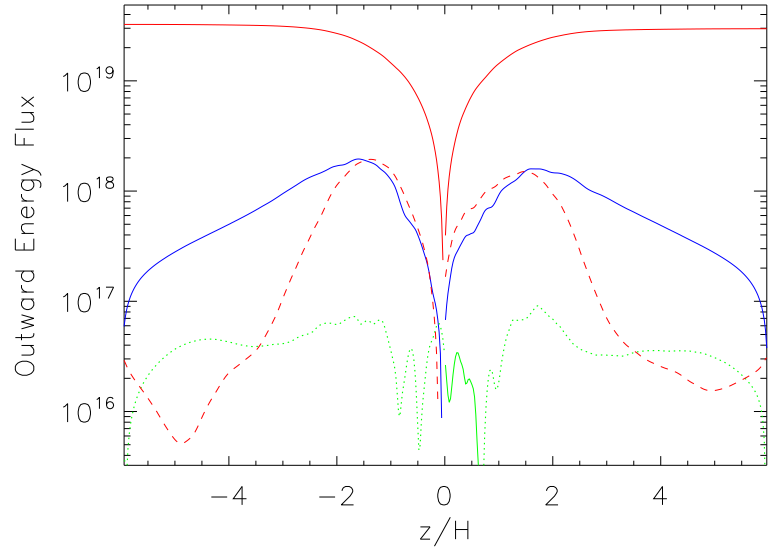,angle=90,width=4.5in}}
\caption{Horizontally- and time-averaged vertical energy fluxes
in erg~cm$^{-2}$~s$^{-1}$.
The solid red curve is diffusive radiation flux, the dashed red curve is
advected radiation flux, the blue curve is Poynting flux, and the green
curves are advected gas energy flux, solid outward, dotted inward.
\label{fig:fluxes}}
\end{figure}

   Diffusive radiation flux dominates the transport of energy up and down
(Fig.~\ref{fig:fluxes}).  In the spirit of our flux-limited diffusion
approximation, for this discussion we label both truly diffusive radiation
transport and free-streaming radiation outside the photosphere as ``diffusive".
Only near $|z| \simeq 2H$ are any other channels at all significant: there,
advected radiative flux and Poynting flux both typically contribute
$\simeq 5\%$ of the total.  At times when the energy content of the
disk is rising, the high-pressure portion of the disk swells, so that
the gas advects both magnetic field and radiation outward.  The peak
in advected radiative flux and Poynting flux occurs at $|z|\simeq 2H$
because this is typically where the gradient in both magnetic field
energy density and radiation energy density is greatest.
Elsewhere in the disk, they and all other
channels (e.g., advected gas energy flux) are very weak.  In fact,
through most of the disk, the time-averaged advected gas energy flux
is {\it inward}, in contrast to all the other fluxes.  Presumably,
this is due to denser regions sinking, while less dense regions,
perhaps with greater radiation energy density, rise. At particular instants,
it is possible for
Poynting flux, advected radiation flux, and advected gas energy to be
comparable to diffusive radiation flux within $\pm 2H$ of the midplane,
but these instances are not long-lasting.

At the top and
bottom boundaries, Poynting flux is the nearest competitor to diffusive
radiation losses, and even it is responsible for at most $\simeq 0.7\%$
(this estimate is made on the basis of the Poynting flux at $|z| \simeq 5.5H$
in order to avoid possible artifacts due to the small artificial resistivity
near the outer boundaries).  If the magnetic field energy in the box were
removed only by Poynting flux, it would take $\simeq 300$~orbits to drain
it completely.   Because the magnetic field energy density can diminish
on much shorter timescales (see Fig.~\ref{fig:energyhistory}), it is clear
that Poynting flux plays only a minor role in determining the saturation
amplitude of the magnetic field.  Because magnetic flux is proportional
to the square root of magnetic energy density and both quantities decrease
outward, it is carried off through the outer boundaries at a
somewhat greater rate (a characteristic loss time of $\simeq 45$~orbits),
but one that is still quite slow compared to the thermal time of
$\simeq 7.5$~orbits.  

    Study of the outgoing energy flux also reveals that it is systematically
slightly smaller than the rate at which work is done on the radial walls.
In terms of the integrated energy up to a given time, this ratio is
$\simeq 0.9$--0.95 up to around 100~orbits, $\simeq 0.98$--0.99 from
$\simeq 110$~orbits until the end of the simulation.  The reason for
this small discrepancy from energy balance is, of course, that the
time-averaged total energy in the volume is rather greater than in
the initial condition, with the energy in the simulation volume building
systematically up to the first big peak at $t=90$~orbits.

     More interestingly, after $t=110$~orbits, the time-integrated
dissipated energy is consistently only about 0.92--0.95 times the
time-integrated outgoing flux.  Because energy is conserved quite
accurately in the code, this discrepancy signals that dissipation
is not the only source of energy for radiation.  Work can be done
on the radiation pressure through bulk compression, raising its
energy non-dissipatively.   Similarly, compressive work done on the
gas can raise its temperature, leading to a higher rate of radiation
emission.  These two mechanisms taken together account for 5--$8\%$
of the total emitted radiation.

\section{Significance}

\subsection{Approximate linear thermal stability analysis}

The crudest possible model for the thermal balance of a disk segment is
\begin{equation}
{dE_{\rm tot} \over dt} = Q(E_{\rm tot}) - E_{\rm tot}/t_{\rm cool},
\end{equation}
where $E_{\rm tot} = e + E$ is the total energy and $Q$ the dissipation,
both per unit area.  Note that in the present context, all quantities
should be regarded as time-averaged over a time-span $\sim t_{\rm cool}$
($\simeq 7$~orbits in this simulation)
because there are sizable random fluctuations in everything over shorter
times.  We have already shown that $Q$ is roughly
$\propto E_{\rm tot}^n$ with $n \simeq 1$, although $n$ declines slowly as
$E_{\rm tot}$ increases.  As we will discuss shortly, $t_{\rm cool}$
in this simulation depends more weakly on $E_{\rm tot}$; we define
$\partial\ln t_{\rm cool}/\partial\ln E_{\rm tot} \equiv s$.

    Linearizing with respect to a perturbation about an equilibrium total
energy per unit area $E_0$, we find
\begin{equation}
{d \delta E_{\rm tot} \over dt} = \frac{\delta E_{\rm tot}}{t_{\rm cool}(E_0)}
           \left(n + s - 1\right)
\end{equation}
because $Q(E_0) = E_0/t_{\rm cool}$ if the initial state was in equilibrium.
Clearly, perturbations grow or decay according to whether $n$ is larger or
smaller than $1-s$.  Over the dynamic range in $E_{\rm tot}$ spanned by our data
(about a factor of four), $n$ declines with increasing energy from $\simeq 1$ to
$\simeq 0.5$. 

     A simple qualitative model explains why $s$, the logarithmic derivative
of $t_{\rm cool}$ with respect to $E_{\rm tot}$, should change from being
large and negative at small radiation content ($e/E \gg 1$) to $\simeq 0$
at intermediate levels ($e \sim E$, the case treated in this simulation), to
$\simeq 1$ in the truly radiation-dominated limit ($e/E \ll 1$).  Because
the gas temperature is tightly coupled to the radiation temperature, the
total thermal content of the disk is the sum of the two energies.  Thus,
the cooling time is
\begin{equation}
t_{\rm cool} \sim \left(\frac{E + e}{F_r}\right) \sim
   \left(\frac{E + e}{E}\right)\left(\frac{H_r \tau}{c}\right),
\end{equation}
where $F_r$ is the radiation flux, $H_r$ is the characteristic thickness of the
trapped radiation and $\tau$ is the typical optical depth photons must overcome
to escape.  With fixed column density, $\tau$ does not change.  When $e \gg E$,
increasing $E$ leads to a diminishing cooling time, as the fraction of
the total energy able to diffuse readily grows while the distance it needs
to travel ($\sim H$) hardly changes.  In this gas-dominated regime, increasing
radiation content makes cooling more rapid, enforcing thermal stability:
in the extreme limit of $e \gg E$, $s \simeq -7$ at fixed column density
because $E \sim H \kappa\Sigma T^4$ and $H \sim e/(\Sigma\Omega^2)$ in
hydrostatic balance.
However, this trend slows when $E \sim e$ (the regime treated here), as
$1 + e/E$ cannot fall below unity.  Formally,
$\partial\ln (1 + e/E)/\partial\ln E_{\rm tot} = -e/E$ if $e$ is
constant, as it more or less is in this simulation.  Our time-averaged ratio
of gas energy to radiation energy is $\simeq 1/2$.
At the same time, in the regime treated here, where $E \sim e$, $H_r$
increases slowly as the radiation begins to take a larger share of support
against gravity.  In fact, if we define the instantaneous radiation scale-height
as $H_r(t) \equiv \int dz |z| E(z,t)/\int dz E(z,t)$, over the
range of $E_{\rm tot}$ sampled by this simulation, $H_r$ is very well fitted by
\begin{equation}
H_r = 0.38H \left(\frac{E_{\rm tot}}{10^{20}\hbox{~erg~cm$^{-2}$}}\right)^{0.4}.
\end{equation}
Thus, this slow rise in $H_r$ as $E_{\rm tot}$ increases very nearly cancels
the reduction in the cooling time due to the increased
share of radiation in the total energy.  The net result is that
in this regime the cooling time
becomes nearly independent of $E$, or $s \simeq 0$.   When $E \gg e$, the ratio
$(E + e)/E \simeq 1$, but $H_r \propto E$, so $s \simeq 1$.  In other words,
in this regime, the cooling time
lengthens with increasing $E$ as higher radiation pressure causes the disk
to expand outward, increasing the total path length photons must travel
before escaping.

   Assembling the pieces, we see that this model suggests $n \simeq 1$
and $s \ll -1$ at low energy content, which is when $e \ll E$.  In
that limit, $n + s -1 \ll -1$, yielding strong thermal stability.  As the
energy content grows, so does $E/e$.  When $E/e$ approaches unity,
$n$ falls slightly below 1, while $s$ climbs toward
zero for the reasons given in the preceding paragraph.  As a result,
$n + s -1 \simeq 0$ in this regime and the system is near
marginal stability.  If $n$ falls further with increasing $E/e$
while $s$ changes only slowly, the system returns to stability.

    Behavior of this sort is exactly what we observe in this simulation.
When the disk has total energy near the bottom of our
range, it readily develops sizable fluctuations, but they do not
grow exponentially with time.  Large positive fluctuations in $E_{\rm tot}$
drive the disk into a regime in which there is weak
negative feedback, so the disk's energy drops back down.  If it falls
far enough, new fluctuations grow.

\subsection{What is $n$?}

     As we have remarked several times already, the volume-integrated dissipation
rate increases with increasing energy in the box, but somewhat more slowly than
linearly.   One possible explanation for this weaker-than-expected dependence
on pressure was initially suggested by Turner et~al. (2003): that in the
radiation-dominated regime, MHD turbulence is not fully coupled to the
pressure because photons can diffuse across a single wavelength of the fastest-growing
MRI mode in less than an orbital time.  When the magnetic field has a significant
toroidal component (in fact, shear usually makes the toroidal component the
largest), the MRI can be compressive.   Photon diffusion can reduce the growth
rate, or even lead to damping, for compressive modes.  It is possible that
lower growth rates lead to lower saturation amplitudes.  Although Turner
et~al. gave an argument showing that radiation-coupling was marginal in
the radiation-dominated regime, this connection between
the photon diffusion time and the MRI in fact applies to all disk states,
radiation-dominated or not.

    If the angular momentum flux through the disk is small compared to
$\dot M r^2 \Omega$, the time-averaged dissipation rate is
$W d\Omega/d\ln r$, where $W$ is the vertically-integrated stress.
Defining $\alpha$ in the usual way as the ratio between $W$ and
the vertically-integrated pressure, the thermal equilibration time
is then $t_{th} = [(\gamma - 1)(d\ln\Omega/d\ln r)\alpha\Omega]^{-1}$, where
$\gamma$ is the effective adiabatic index.
The distance photons can diffuse in any given time
$t$ is $l_D(t) = [ct/(3\kappa\rho)]^{1/2}$.  In particular, in thermal
equilibrium, $l_D = H$ for $t=t_{th}$, so
\begin{equation}
H = \left[\left(\gamma - 1\right)\left(d\ln\Omega/d\ln r\right)\alpha\right]^{-1/2} 
   \left[\frac{c}{3\kappa\rho\Omega}\right]^{1/2}.
\end{equation}
On the other hand, the growth time of the fastest-growing MRI mode is always
$\simeq \Omega^{-1}$, so we see that
\begin{equation}
H = \left[\left(\gamma - 1\right)\left(d\ln\Omega/d\ln r\right)\alpha\right]^{-1/2}
     l_D (t_{gr}).
\end{equation}

     Now the wavelength of the fastest-growing MRI mode is $2\pi v_{Az}/\Omega$,
where $v_{Az}$ is the Alfv\'en speed taking into account only the vertical
component of the magnetic field.  We then find that the ratio of that
wavelength to the distance photons can diffuse in its growth time is
\begin{equation}
\frac{\lambda_Z}{l_D (t_{gr})} = \frac{2\pi c_t^2}{H\Omega}
        \left[\gamma\left(\gamma - 1\right)\left(d\ln\Omega/d\ln r\right)\right]^{-1/2}
       \frac{\langle|B_z|\rangle}{\langle B_r B_\phi\rangle^{1/2}}.
\end{equation}
Because $H\Omega \simeq c_t^2$ and
$\langle|B_z|\rangle/\langle B_r B_\phi\rangle^{1/2} \simeq 0.3$--0.5, we
therefore find that $\lambda_Z/l_D(t_{gr})
\simeq 1$ in all circumstances, radiation-dominated or not.  In other words,
dynamical coupling between radiation pressure and the fastest-growing MRI
mode is always marginal.

      We speculate that the partial inability of the dissipation rate (and
stress) to scale with the radiation pressure is due to this imperfect dynamical
coupling between the basic driver of MHD turbulence (and therefore stress)
and the radiation pressure.   One way to estimate the magnitude of this
effect is to use the formalism of Blaes \& Socrates (2001,2003) to compute
the MRI growth rate under these circumstances.  In the limit of perfect
thermal coupling between gas and radiation ($\omega_a \rightarrow \infty$
in their notation), we find that in the conditions of this simulation
the maximum local linear growth rate can be moderately suppressed (generally
speaking, by $\sim 30\%$).  As already discussed, the degree of suppression
increases with growing $l_D/\lambda_z$.  In addition, because compressibility
grows in importance with increasing $|B_\phi/B_z|$, larger values of this
ratio also reduce the growth rate when photon diffusion is important; in
this simulation, typical values of $|B_\phi/B_z|$ in the disk body were
3--5.

\subsection{Implications for hydrostatic balance and stability at higher
radiation gas pressure ratios}

    There is a critical dissipation scale for radiation pressure-supported,
geometrically-thin disks: $c\Omega^2/\kappa$.   When the dissipation per unit
volume matches this at all altitudes below the point of observation, the upward
radiation force exactly matches the downward component of gravity; if the
net force from gas pressure gradients and magnetic fields is upward, the
disk matter must move away from the midplane.
Measured in terms of this critical value, the mean
dissipation rate within a few scale-heights of the midplane during times of
high total energy in this simulation was $\simeq 0.7$; during times of low
total energy, it was only about half as large, $\simeq 0.35$.  In other words,
there are episodes as long as $\sim 5t_{\rm cool}$ during which most of the
disk mass derived $70\%$ of its support against vertical gravity from radiation
forces, but for the remainder of the simulation the fractional support
due to radiation was smaller.  At higher altitudes, the magnetic pressure
gradient replaces radiation as the primary source of vertical force at all
times.

These mean values cover a great deal of fluctuation, however.   Even after
averaging over a cooling time (7 orbits) sampled every 0.01 orbits, there
are sizable fluctuations in the horizontally-averaged dissipation rate within
$2H$ of the midplane, where most of the dissipation occurs.  In that region,
the instantaneous horizontally-averaged dissipation rate typically varies
by factors of 3--4, with the scale of
variation often as little as $\sim H/4$.  If one considers fluctuations
within individual planes relative to the instantaneous mean for that plane,
the rms dispersion is typically 2--4 times the mean, but can be as much as
5--10 times the mean.  Not surprisingly, dissipation features within a single
plane tend to form radially sheared stripes stretched in the azimuthal
direction.

Because the fluctuations are so large, it is also useful to consider
the question of fractional radiation support in terms of the fraction
of the time that the dissipation rate matches or exceeds the critical
value.  In the conditions of this simulation, this duty cycle was
$\simeq 0.2$ within $\pm 2H$ of the midplane.  Were it to rise above
$\simeq 0.5$, one could no longer expect hydrostatic
equilibrium in the central part of the disk because radiation
force by itself would exceed gravity more often than not.

As a plausible guess, we might estimate that this duty cycle would
be linearly proportional to the time-average of the dissipation rate.
Thus, an increase in the mean dissipation rate per unit volume by a factor
of 2--3 relative to what was seen in this simulation would lead to difficulties
even in establishing hydrostatic equilibrium.  Because the vertically-integrated
dissipation rate is $\propto r^{-3}$, whereas the disk thickness changes
only slightly with radius when radiation dominates its vertical support,
such an increase might be expected at radii only 0.7--0.8 times the one
at which this simulation was conducted (i.e., at $r \simeq 110$--$120r_g$).
In addition, if we employ the toy-model of \S~4.1, we might predict
that the cooling time would begin
to grow with increasing $E$, perhaps to the point that $s\simeq 1$.  We
know little about the dependence of the dissipation rate on $E$ in this
regime; however, if $n > 0$ and $s \simeq 1$, it might be difficult to
avoid a thermal instability similar to the one envisaged by Shakura
\& Sunyaev (1976).

On the other hand, it is also possible that photon bubbles (Gammie 1998,
Begelman 2001) might sufficiently
enhance the radiation transport rate relative to conventional diffusion through
a more homogeneous substrate that thermal instability could be quenched.
Put another way, photon bubbles might keep $s$ well below unity.
As discussed in the companion paper (Blaes et al. 2007), we do not find any
strong evidence for
photon bubbles in the current simulation, even though linear theory predicts
they should grow rapidly in the upper layers of this simulation, and we have
adequate resolution to see the fastest-growing modes.   Nonetheless, it
remains possible that they could be more important at higher levels of
radiation pressure support.  This is because photon bubble growth rates
are proportional to the radiative flux, so that they may be able
to form and reform rapidly even in the presence of competing time-dependent
phenomena.  It may be challenging to find them in a simulation, however, as
the expected characteristic wavelength of these instabilities is the gas
pressure scale height, which can be very small in a radiation dominated
disk.

\subsection{A new high radiation pressure equilibrium?}

    These results may be combined to form a new approximate equilibrium for
the radiation-dominated state.  Such an equilibrium might be a suitable
initial condition for future simulations exploring higher ratios of
radiation to gas pressure than found here.

     Let us begin with the supposition that the dissipation rate per unit
mass continues to be $\propto m^{-1/2}$, independent of the ratio of
radiation to gas energy.  If, in addition, the dissipation rate per unit
volume is fixed to the critical value for pure radiation support in
hydrostatic balance, then we have
\begin{equation}\label{eq:critdiss}
Q/\rho = \frac{c\Omega^2}{\kappa\rho} = A m^{-1/2},
\end{equation}
for some proportionality constant $A$.  However, since $dm/dz = -\rho$,
Equation~\ref{eq:critdiss} leads to the differential equation
\begin{equation}
{dm \over dz} = - \frac{c\Omega^2}{\kappa A}m^{1/2},
\end{equation}
which has the solution
\begin{equation}
m(z) = (1/2)\Sigma\left(1 - z/z_{\rm max}\right)^2,
\end{equation}
where $z_{\rm max} = 2\kappa A/(c\Omega^2)$.   The proportionality constant
$A$ can be determined in terms of the total mass accretion rate; the result
is that $z_{\rm max} = 3\dot M/(8\pi c)$, just as in the usual Shakura
\& Sunyaev (1973) radiation-dominated equilibrium (neglecting correction
factors due to the net flux of angular momentum through the disk and
relativistic effects).  Unlike that equilibrum, however, in which the
density was supposed to be constant, here
\begin{equation}
\rho(z) = \left(\Sigma/z_{\rm max}\right)\left(1 - z/z_{\rm max}\right).
\end{equation}

     This equilibrium neglects any magnetic support.   From what we have
learned from this simulation and our previous one, we now know that the
upper layers of the disk are generally magnetically-supported.  In reality,
then, it is likely that, rather than going to zero at $z_{\rm max}$, the
gas density actually stretches beyond $z_{\rm max}$ and tapers more gradually
to zero.

\section{Summary}

   In this paper we have explored the interplay of MHD turbulence stirred
by the magneto-rotational instability with radiation forces in a regime
in which the radiation pressure is comparable to the gas pressure.

    Several results stand out:

\noindent 1) As in our previous simulation of a gas pressure-dominated
shearing box, although a mix of radiation and gas pressure accounts for
most of the material's support against vertical gravity near the disk
plane, the outer layers of the disk are held up primarily by magnetic
pressure gradients.  More than half of the disk's volume inside the
photosphere is magnetically-dominated, although these layers account
for a much smaller fraction of its total mass.

\noindent 2) The energy content of the disk undergoes large (factor of
several) excursions, with typical episodes lasting $\simeq 4$--6 cooling
times.  These do not lead to runaway, despite the fact that radiation
pressure can be several times as great as gas pressure for several cooling
times.  Both the large fluctuations and the absence of instability can
be explained qualitatively by a combination of a slower-than-linear
rise of dissipation rate with energy content and, when gas and
radiation energies are comparable, the insensitivity of the cooling time
to changes in total energy.

\noindent 3) We extend the domain of applicability of our previous result,
that the dissipation is correlated with the current density, rising in the mean
$\propto |J|^{1 + \epsilon}$, where $\epsilon \ll 1$, but with a very
sizable scatter around this mean relationship.  The time-average dissipation
rate per unit mass appears also to depend on optical depth, scaling
$\propto m^{-1/2}$ when considering time- and horizontally-averaged values.
On the other hand, these trends apply only to averages:  the
dissipation is highly intermittent and time-variable, with a small fraction
of the total cells ($\sim 10^{-3}$) accounting for 3/4 of the total at
any one time.  In addition, a small part of the total radiated energy (5--8\%)
has its origin in non-dissipative processes, compressive work done on
either the gas or the radiation.

We would like to thank Jim Stone for useful suggestions, and also
for developing the simulation code we used.
This work was supported in part by NSF Grants AST-0307657 (OB) and AST-0507455
(JHK).  Numerical  calculations were carried out on the SX8 at the Yukawa
Institute for Theoretical Physics of Kyoto University and the VPP5000 at
the Center for Computational Astrophysics of the National Astronomical
Observatory of Japan.

\end{document}